\documentclass[a4paper,11pt]{article}
\usepackage{graphicx}
\usepackage{color}
\usepackage{ulem}
\usepackage{enumerate}
\usepackage[colorlinks=true,linkcolor=red,citecolor=blue,urlcolor=blue,bookmarks]{hyperref}
\usepackage{amssymb}
\usepackage{subfigure}
\def \be {\begin{equation}}
\def \ee {\end{equation}}
\def \ba {\begin{array}}
\def \ea {\end{array}}
\def \bea {\begin{eqnarray}}
\def \eea {\end{eqnarray}}

\def \ble {\begin{widetext}\begin{equation}}
\def \ele {\end{equation}\end{widetext}}
\def \blea {\begin{widetext}\begin{eqnarray}}
\def \elea {\end{eqnarray}\end{widetext}}

\def \nn {\nonumber}

\def \trans {\mathcal{T}^{\psi|\phi}}
\def \transA {\mathcal{T}_A^{\psi|\phi}}

\def \blea {\begin{widetext}\begin{eqnarray}}
\def \elea {\end{eqnarray}\end{widetext}}
\def \mO {\mathcal{O}}

\begin{document}
\title{Non-Hermitian spacetime and generalized thermofield double formalism}
\author{Wu-zhong Guo\footnote{wuzhong@hust.edu.cn}~, Tao Liu\footnote{m202270240@hust.edu.cn}}

\date{}
\maketitle

\vspace{-10mm}
\begin{center}
{\it School of Physics, Huazhong University of Science and Technology,\\
Luoyu Road 1037, Wuhan, Hubei
430074, China
\vspace{1mm}
}
\vspace{10mm}
\end{center}

\begin{abstract}
In this paper, we explore the non-Hermitian transition matrix and its gravity dual. States in quantum field theories or gravity theories are typically prepared using Euclidean path integrals. We demonstrate that it is both natural and necessary to introduce non-Hermitian transitions to describe the state when employing different inner products in Euclidean quantum field theories. Transition matrices that are $\eta$-pseudo-Hermitian, with $\eta$ being positive-definite, play the same role as density matrices, where the operator $\eta$ is closely related to the definition of the inner product. Moreover, there exists a one-to-one correspondence between these transition matrices and density matrices. In the context of AdS/CFT correspondence, the Euclidean path integral in the boundary field theory can be translated to the bulk gravitational path integral. We provide an overview of the construction and interpretation of non-Hermitian spacetime. Specifically, we demonstrate the crucial role of the non-Hermitian transition matrix in realizing the thermofield concept in general cases and in understanding the gravity states dual to the eternal black hole. In this context, the pseudoentropy of the transition matrix can also be interpreted as black hole entropy. Finally, we highlight the strong subadditivity property of pseudoentropy, and the connection between non-Hermitian transition matrices and complex metrics.
\end{abstract}

\tableofcontents

\section{Introduction}

The wave function describes the states of a given quantum system. In a finite-dimensional Hilbert space, it is possible to solve the wave function and express it using a complete basis of the system. However, in quantum field theories (QFTs), it is generally not possible to obtain such an expression even if a basis is chosen. Instead, the Euclidean path integral is typically used to represent the wave function of the states in QFTs or quantum gravity \cite{Hartle:1983ai}\cite{Hawking:1983hj}.

The state properties can be detected by investigating local or non-local observables. One of the most useful quantities is the entanglement entropy (EE) \cite{Calabrese:2004eu}\cite{Calabrese:2009qy}, which is closely related to the basic properties of QFTs \cite{Srednicki:1993im}-\cite{Balakrishnan:2017bjg}. In the context of AdS/CFT, at the leading order of $G$, the EE is given by the Ryu-Takayanagi (RT) formula \cite{Ryu:2006bv} and its generalization, the Hubeny-Rangamani-Takayanagi (HRT) formula \cite{Hubeny:2007xt}. Understanding the entanglement structure of wave functions in QFTs and the gravity dual of these wave functions is an important task. The RT formula is a fundamental tool for comprehending the properties of both classical and quantum spacetimes \cite{VanRaamsdonk:2010pw}-\cite{Almheiri:2019psf}.

Suppose one has prepared the ket state $|\Psi\rangle$. It is straightforward to represent the density matrix $|\Psi\rangle \langle \Psi|$ by defining the bra state $\langle \Psi|$ as the Hermitian conjugate of the ket state. In the usual approach in flat spacetime, one takes the fields on the timeslice $x_0=0$ as the basis. The timeslice in Euclidean QFTs can be naturally obtained by Wick rotation of Lorentzian QFTs. Generally, density matrices, recognized as Hermitian operators, can also be used to describe the states of the system. However, the Hermitian conjugation of ket states or operators depends on the quantization method employed in the theory. In Euclidean QFTs, any coordinates can be taken as time. Different choices of the time coordinate correspond to different quantizations of the theory. More generally, one could choose any codimension-1 surface to quantize Euclidean QFTs \cite{Simmons-Duffin:2016gjk}. Therefore, in Euclidean QFTs, different quantization methods will lead to different definitions of the inner product, resulting in different Hermitian conjugation operations on bra states and operators. In this paper, we will explore how to represent the states by Euclidean path integrals when different quantization methods are used.

Below, we will demonstrate that, to describe the system in field theory or gravity theory in general cases, it is both natural and necessary to introduce the non-Hermitian transition matrix
\bea
\mathcal{T}=|\psi\rangle \langle\phi|,
\eea
with $\langle \phi| \ne |\psi\rangle^\dagger$.
We provide explicit examples to illustrate the emergence of the non-Hermitian transition matrix by selecting different quantization methods in Euclidean QFTs.

Recently, the transition matrix was introduced in \cite{Nakata:2020luh} with the primary motivation to define the pseudoentropy (PE), which can be considered a generalization of entanglement entropy (EE). In the context of AdS/CFT, if the transition matrix is dual to bulk geometry, the PE can also be evaluated using the RT formula, provided the transition matrix corresponds to some geometry at the leading order of $G$. In \cite{Guo:2022jzs}, the authors use the concept of pseudo-Hermiticity to characterize the real-valued condition of PE. However, the physical meaning of this condition is still not well understood.

In this paper, we will provide a physical explanation for the pseudo-Hermitian conditions discussed in \cite{Guo:2022jzs}. Specifically, we will show that if one employs a different inner product for a given quantum system, it becomes natural to describe the system using an $\eta$-pseudo-Hermitian transition matrix, which plays the same role as the density matrix. The operator $\eta$ is positive-definite and is closely related to the inner product. Furthermore, we will demonstrate that a duality exists between the Hermitian density matrix and the non-Hermitian transition matrix, implying a one-to-one correspondence between their observables.

The pseudo-Hermitian transition matrix resembles the recent interest in PT-symmetric or pseudo-Hermitian quantum mechanics, which focuses on the necessary and sufficient conditions for the reality of the Hamiltonian spectrum \cite{Bender:2007nj,Mostafazadeh:2008pw,Ashida:2020dkc}. In fact, we will show below that the pseudo-Hermitian transition matrix can be well-understood within the framework of pseudo-Hermitian quantum mechanics\cite{Mostafazadeh:2001jk,Mostafazadeh:2001nr,Mostafazadeh:2004mx}.
To describe the quantum states of gravity, we must also take into account the concept of non-Hermitian spacetime. This concept can be better articulated within the framework of the AdS/CFT correspondence. The path integral representation of states in boundary QFT can be translated to the bulk gravitational path integral. In the semiclassical limit $G \to 0$, the non-Hermitian property reflects the asymmetry of the metric along certain surfaces. We will use some examples to show how non-Hermitian spacetime offers further insight into the properties of gravity. As an application, we construct a non-Hermitian transition matrix based on the thermofield double (TFD) state. We will show that the transition matrix can also be used to realize the thermofield double idea in more general situations. Furthermore, we point out that the transition matrix can be dual to the eternal black hole, just like the TFD density matrix.

The paper is organized as follows. In Section \ref{section_second}, we discuss the relationship between states, density matrices, and transition matrices. Specifically, we show the emergence of the transition matrix when employing different quantization methods in Euclidean QFTs. We also demonstrate the relation between $\eta$-pseudo-Hermiticity and the inner product defined in association with the metric operator $\eta$. In Section \ref{section_gravity_dual}, we present the gravity dual of the non-Hermitian transition matrix. We provide examples to illustrate the general properties of non-Hermitian spacetime. Additionally, we construct the non-Hermitian transition matrix (\ref{generalized_TFD}) based on the thermofield double state, showing that (\ref{generalized_TFD}) can equivalently realize the thermofield double idea. The gravity dual of (\ref{generalized_TFD}) is proposed to be related to the eternal black hole with asymmetric partition. Section \ref{section_discussion} includes the conclusions and discussions. We highlight some interesting problems arising from our work. The Appendix contains a review of the metric operator for defining inner products in finite-dimensional Hilbert spaces, along with details of the calculations in Section \ref{section_gravity_dual}.

\section{States, density matrix and transition matrix}\label{section_second}

The information of a given quantum system is encoded in the wave function $|\Psi\rangle$. Alternatively, in standard quantum mechanics, we can also use the density matrix $\rho_\Psi := |\Psi\rangle \langle \Psi|$, which satisfies
\bea\label{densitymatrix}
\langle \Psi| a |\Psi\rangle = tr(a \rho_\Psi), 
\eea
where $a$ represents arbitrary observables. Generally, $\rho_\Psi$ is a Hermitian operator. However, it is essential to realize that the operation of Hermitian conjugation depends on the definition of the inner product and the quantization method employed in the theory.

Suppose one has a Hilbert space with a well-defined inner product $\langle \cdot | \cdot \rangle$. It is possible to introduce a new inner product associated with the \textit{metric operator} $\eta$:
\bea\label{innerproductrelation}
\langle \cdot | \cdot \rangle_\eta := \langle \cdot | \eta \cdot \rangle.
\eea
The metric operator is positive and invertible. In a finite-dimensional Hilbert space, there is a one-to-one correspondence between the metric operator and the inner product \cite{Mostafazadeh:2004mx,Mostafazadeh:2008pw}. The metric operator can also be constructed using the bases, see Appendix \ref{appendix_metric} for details.

 In the following, we will first show different quantization methods in Euclidean QFTs and the definition of Hermitian conjugation. We will also argue that it is natural to consider a special class of non-Hermitian transition matrices to describe the states of a given system.



\subsection{Different quantization in Euclidean QFTs}

For Lorentzian QFTs the energy and momentum operators $(H,\bf{\vec{P}})$ are Hermitian. If an operator $\mathcal{O}(0,0)$ is Hermitian, then \bea
\mathcal{O}(t,\vec{x}):=e^{iH t-i \vec{x}\cdot \bf{\vec{P}}}\mathcal{O}(0,0)e^{-iH t+i \vec{x}\cdot \bf{\vec{P}}}, \nn
\eea 
is also Hermitian. While in Euclidean QFTs the situation is different. In usual approach one could obtain the Euclidean theory by Wick rotation $t\to -i t_E$, where $t_E$ is the Euclidean time. For the operator in Euclidean theory $\mathcal{O}_E(\tau_E,\vec{x}):= \mathcal{O}(it,\vec{x})$, we would have the Hermitian conjugation defined by $\mathcal{O}_E(\tau_E,\vec{x})^\dagger =\mathcal{O}(-\tau_E,\vec{x})$. Thus the local operator $\mathcal{O}_E(\tau_E,\vec{x})$ with $\tau_E\ne 0$ is no longer Hermitian operator.

However, in Euclidean QFT the quantization approach can differ based on which direction is considered as ``time''. Selecting different time coordinate results in the construction of distinct Hilbert spaces for the theory, consequently yielding different Hermitian conjugation operations, see, e.g., \cite{Simmons-Duffin:2016gjk}. 

Let's consider a quantum field theory defined on Euclidean spacetime $\mathbb{R}^d$ with coordinates ${x_i}$ ($i=0,...,d-1$). Then we would have the operator $P^i$ ($i=0,...,d-1$) which generates translations in the $x_i$ coordinate. We have the flexibility to designate any direction as the ``time'' coordinate and then proceed to quantize the theory on a constant time slice. The choice of this ``time'' direction determines the Hilbert space.

For instance, let's take $x_0$ as the time coordinate. In this case, the Hamiltonian $P^0$, which drives time evolution, is a Hermitian operator. For any local operator $\mathcal{O}(x_0)$, we would expect $\mO(x_0)=e^{x_0 P^0}\mO(0)e^{-x_0 P^0}$, which satisfies $\mO(x_0)^\dagger=\mO(-x_0)$ for the Hermitian operator $\mO(0)$. If we have a different time direction, say $x_1$,  then the operator $P^1$ is Hermitian, considered to be Hamiltonian. Then you will have different version of Hermitian conjugation. In Euclidean QFTs Hermitian conjugation depends on how to quantize the theory. 
The definition of Hermitian conjugation is also associated with the inner product definition in Euclidean QFTs like the finite dimension example shown in Appendix.\ref{appendix_metric}.

\subsubsection{Quantization with different time coordinate}\label{section_different_time}
In QFTs we usually use the Euclidean path integral to prepare the states. This entails selecting a time direction. The states within the Hilbert space are then constructed by acting local operators on the vacuum state $|0\rangle$, which can be prepared through the Euclidean path integral with inserting operators. In this section, we will demonstrate how one should use non-Hermitian transition matrices to describe the state when choosing different time coordinates in Euclidean QFTs.

\begin{figure}
\centering\includegraphics[width=10cm]{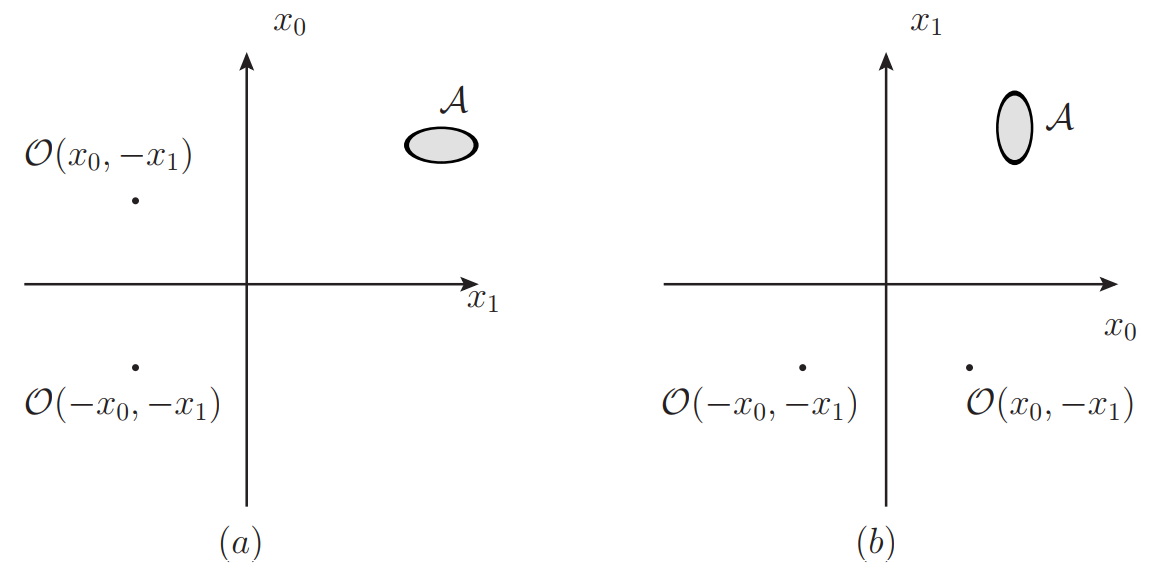}
\caption{Path integral representation of the density matrix $\rho$ and $\mathcal{T}$. Here, $\mathcal{A}$ represents the probe operator located within the region shaded in gray. (a) illustrates $\rho$, with $x_0$ chosen as the time coordinate. (b) illustrates $\mathcal{T}$, with $x_1$ serving as the time coordinate.  }\label{fig1}
\end{figure}

To simplify the notation let us consider two-dimensional QFTs with coordinate $(x_0,x_1)$.  An example is the state $|\psi\rangle= \mathcal{O}(-x_0,-x_1)|0\rangle$ with $x_0,x_1>0$. Let us take $x_0$ as the time coordinate. One could choose a basis $|\phi(x_0=0,x_1)\rangle$, where $\phi(x_0=0,x_1)$ denotes the fields on the time slice $x_0=0$. $\langle \phi(x_0=0, x_1)|\psi\rangle$ can be expressed as a path integral over the lower half of Euclidean space with insertion of the operator $\mathcal{O}(-x_0,-x_1)$. The unnormalized density matrix 
\bea\label{density_matrix_time}
\rho=|\psi\rangle\langle \psi|=\mathcal{O}(-x_0,-x_1)|0\rangle \langle 0| \mathcal{O}(x_0,-x_1),
\eea 
is a Hermitian operator, which can also expressed as path integral over the whole complex plane with insertion of the operators $\mathcal{O}(-x_0,-x_1)$ and $\mathcal{O}(x_0,-x_1)$, see Fig.\ref{fig1} for the illustration.

Now, let's consider the same state but with $x_1$ as the time direction and choose the basis $|\phi(x_0,x_1=0)\rangle$. The Hilbert space is different from the one we considered above. As we can see from Fig.\ref{fig1}, the state is more properly described by the unnormalized transition matrix:
\bea\label{transition_matrix_time}
\mathcal{T}=\mathcal{O}(x_0,-x_1)\mathcal{O}(-x_0,-x_1)|0\rangle\langle 0|,
\eea
which is obviously non-Hermitian. Thus, it is natural for the non-Hermitian transition matrix to appear when using different quantization methods.

We will argue that $\rho$ and $\mathcal{T}$ can effectively capture the same system state. Evidently, we find $tr(\rho) = tr(\mathcal{T}) = \langle 0| \mathcal{O}(x_0,-x_1)\mathcal{O}(-x_0,-x_1)|0\rangle$\footnote{More precisely, the two traces should differ because the Hilbert spaces are distinct. Although we use the same notation here, it's important to bear in mind this difference.}. Moreover, for any local probe operator $\mathcal{A}$ (refer to Fig.\ref{fig1}), we have $tr(\rho\mathcal{A}) = tr(\mathcal{T}\mathcal{A})$. Consequently, physical observables yield indistinguishable outcomes under either description. This consistency is expected as we're employing two distinct quantization methods for the same theory. Both $\rho$ and $\mathcal{T}$ encapsulate the system's information. Thus, it's natural to introduce and explore the presence of non-Hermitian transition matrices in Euclidean QFTs.

\subsubsection{Quantization on different timeslice}\label{section_different_slice}

\begin{figure}
\centering\includegraphics[width=10cm]{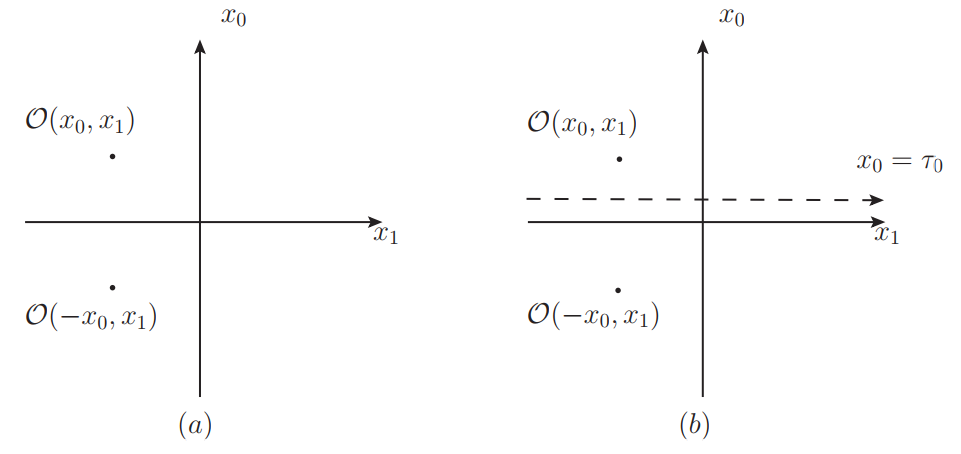}
\caption{Quantization of Euclidean QFTs on different timeslice, $x_0$ is taken to be the time coordinate. (a) The field on $x_0=0$ as the basis. (b) The field on $x_0=\tau_0$ as the basis. }\label{differentslice}
\end{figure}

Even if we have chosen a time coordinate, there is still freedom to choose different bases. Consider $x_0$ as the time coordinate. One could select $|\phi(x_0=0, x_1)\rangle$ and $|\phi(x_0=\tau_0, x_1)\rangle$ as two different bases, as illustrated in Fig.\ref{differentslice}. We will demonstrate below that this corresponds to defining different inner products for the theory. Furthermore, the inner product is directly related to a positive operator $\eta$, which is referred to as the metric operator in a finite-dimensional Hilbert space. 

The usual approach is to choose the basis $|\phi(x_0=0, x_1)\rangle$. As we have shown above, in this case, the Hermitian conjugation is given by $\mathcal{O}(x_0, x_1)^\dagger = \mathcal{O}(-x_0, x_1)$. We can construct a set $\mathcal{H}_{x_0<0} := {\mathcal{A}(x_0<0)|0\rangle}$, where $\mathcal{A}(x_0<0)$ represents arbitrary operators located in the lower half-plane $x_0<0$. The set $\mathcal{H}{x_0<0}$ is dense in the Hilbert space of the theory. For any states $|\phi\rangle, |\psi\rangle \in \mathcal{H}_{x_0<0}$, we can define the inner product $\langle \psi |\phi \rangle$. For more details on constructing the Hilbert space of Euclidean QFTs via Euclidean correlation functions, one could refer to \cite{Osterwalder:1973dx}\cite{Osterwalder:1974tc}. For the special case where $|\phi\rangle = |\psi\rangle = \mathcal{O}(-x_0, x_1)|0\rangle$, the inner product becomes:
\bea
\langle \phi|\phi\rangle=\langle 0| \mathcal{O}(x_0,x_1)\mathcal{O}(-x_0,x_1)|0\rangle\ge 0,
\eea
which is known as refletion positivity. In this case the state of the system shown in Fig.\ref{differentslice}(a) represents the density matrix 
\bea\label{rho_slice}
\rho=\mathcal{O}(-x_0,x_1)|0\rangle \langle 0|\mathcal{O}(x_0,x_1) ,
\eea
which is Hermitian operator.

When using a different basis, $|\phi(x_0=\tau_0, x_1)\rangle$, we employ a different quantization method. The inner product and Hermitian conjugation differ from the case discussed above. In this basis, the state $\mathcal{O}(-x_0, x_1)|0\rangle$ shown in Fig.\ref{differentslice}(b) is prepared by a Euclidean path integral over the region $x_0 < \tau_0$ with the insertion of the operator $\mathcal{O}(-x_0, x_1)$.

Consider an arbitrary state $|\phi\rangle = \mathcal{O}(-x_0, x_1)|0\rangle$ with $-x_0 < \tau_0$. Its Hermitian conjugate is given by
\bea
~_{\tau_0}\langle \phi|:=\left(\mathcal{O}(-x_0, x_1)|0\rangle\right)^\dagger = \langle 0| \mathcal{O}(x_0 + 2\tau_0, x_1),
\eea
which differs from the state $\langle 0|\mathcal{O}(x_0, x_1)$. The subscript $\tau_0$ denotes the Hermitian conjugation with respect to $\tau_0$. Therefore, the operator $\rho$ as defined in (\ref{rho_slice}) is no longer a Hermitian operator in this new basis.

This might seem a bit strange at first. In fact, we have defined a different inner product. The new inner product is related to the previous one by a metric operator $\eta$, similar to the finite-dimensional example. In Appendix \ref{appendix_metric}, we have shown that there are close relationships among the inner product, Hermitian conjugation, and the metric operator $\eta$ in a finite-dimensional Hilbert space. It is also possible to construct the metric operator $\eta$ using the basis.

With the basis $|\phi(x_0 = \tau_0, x_1)\rangle$, the set of states $\mathcal{H}_{x_0 < \tau_0} := {\mathcal{A}(x_0 < \tau_0)|0\rangle}$ is dense in the Hilbert space. Consider the states $|\phi\rangle = \mathcal{O}(x_0, x_1)|0\rangle$ and $|\psi\rangle = \mathcal{O}(x'_0, x'_1)|0\rangle$ with $x_0, x'_0 < \tau_0$. The inner product of these two states is given by
\bea\label{inner_product_new}
\langle \psi| \phi\rangle{\tau_0} = \langle 0|\mathcal{O}(x'_0 + 2\tau_0, x'_1)\mathcal{O}(x_0, x_1)|0\rangle.
\eea
For the case $|\psi\rangle = |\phi\rangle$, we get the norm of the state $|\phi\rangle$,
\bea
\langle \phi|\phi\rangle{\tau_0} = \langle 0|\mathcal{O}(x_0 + 2\tau_0, x_1)\mathcal{O}(x_0, x_1)|0\rangle,
\eea
which demonstrates reflection positivity along the timeslice $x_0 = \tau_0$.

The new inner product (\ref{inner_product_new}) is associated with the old inner product by the following relation
\bea\label{inner_relation_tau0}
\langle \psi| \phi\rangle_{\tau_0}=\langle \psi| \eta_{\tau_0} |\phi\rangle,
\eea
where $\eta_{\tau_0}:= e^{-2\tau_0 H}$. It is obvious $\eta_{\tau_0}$ is positive and invertible operator, which can be taken as the metric operator.

In finite-dimensional Hilbert spaces, it is shown that the metric operator corresponds one-to-one with the inner product. The metric operator is given by Eq.(\ref{metric_finite_dimension}) in Appendix.\ref{appendix_metric}. In fact, we observe a similar relationship in Euclidean QFTs. Using the basis $|\phi(x_0 = 0, x_1)\rangle$, the state $|\psi\rangle$ can be expressed as
\bea\label{complete_0}
\int D\phi(x_1) |\phi(x_0 = 0, x_1)\rangle\langle \phi(x_0 = 0, x_1)|\psi\rangle = |\psi\rangle,
\eea
where $\int D\phi(x_1)$ denotes the path integral over the fields on the timeslice $x_0 = 0$. If we use the inner product (\ref{inner_product_new}), the basis $|\phi(x_0 = 0, x_1)\rangle$ does not satisfy the completeness relation (\ref{complete_0}) since the Hermitian conjugate of $|\phi(x_0 = 0, x_1)\rangle$ is given by $~_{\tau_0}\langle \phi(0, x_1)| = \langle \phi(2\tau_0, x_1)|$. According to (\ref{metric_finite_dimension}), the metric operator $\eta_{\tau_0}$ is given by
\bea
&&\eta_{\tau_0}|\psi\rangle = \int D\phi(x_1) |\phi(x_0 = 0, x_1)\rangle\langle \phi(2\tau_0, x_1)|\psi\rangle \nn \\
&&= \int D\phi(x_1) |\phi(x_0 = 0, x_1)\rangle\langle \phi(x_0 = 0, x_1)|e^{-2\tau_0 H}|\psi\rangle \nn \\
&&= e^{-2 \tau_0 H}|\psi\rangle,
\eea
which is consistent with the previous result (\ref{inner_relation_tau0}).

Now let us consider the states in the two different quantizations, as shown in Fig.\ref{differentslice}. The states can be expressed as the operator $\mathcal{O}(x_0, x_1)|0\rangle \langle 0| \mathcal{O}(-x_0, x_1)$. These states can be prepared using Euclidean path integrals. For the case shown in Fig.\ref{differentslice}(a), the operator is a Hermitian density matrix. However, for the case shown in Fig.\ref{differentslice}(b), this operator should be understood as a non-Hermitian transition matrix.
In this example, we also see the natural necessity of introducing the non-Hermitian transition matrix to describe the states if one quantizes the theory on different timeslices. In the following, we will further explore the relationship between these two descriptions.

\subsection{Hermitian and non-Hermitian duality}\label{section_hermitian_nonhermitian}

In previous discussion we see Hermitian conjugation is directly related to the inner product.  In Euclidean QFTs we use example to show quantization on different timeslice $\tau_0$ leads to different definition of inner product associated with the metric operator $\eta_{\tau_0}$. We expect in more general case there exists one-to-one correspondence between the postive operator $\eta$ and inner product, just like the finite dimensional Hilbert space. We denote the Hilbert space associated with $\eta$ as $\mathcal{H}_\eta$. 

Given a Hilbert spaces $\mathcal{H}$ and define the inner product of two states $|\psi\rangle$ and $|\phi\rangle$ by $\langle \psi| \phi\rangle$. For a linear operator $a$  
one could introduce the Hermitian conjugation $a^\dagger$ by 
\bea
\langle \psi| a^\dagger \phi\rangle =\langle a\psi|\phi\rangle,
\eea
where $|\psi\rangle$ and $|\phi\rangle$ are two arbitrary states in $\mathcal{H}$.

For two states $|\phi\rangle$ and $\psi\rangle$ with the new inner product $\langle \psi|\phi\rangle_\eta$ defined as (\ref{innerproductrelation}), the Hermitian conjugation of a given operator $A$ is defined as
\bea
\langle \psi|A \phi\rangle_\eta=\langle \psi A^\dagger| \phi\rangle_\eta.
\eea
By using the relation (\ref{innerproductrelation}) we find
\bea
\langle \psi|\eta A| \phi\rangle=\langle \psi| A^\dagger \eta |\phi\rangle.
\eea
The above relation is correct for arbitrary $|\phi\rangle$ and $|\psi\rangle$. Thus we would have
\bea
A^\dagger=\eta A \eta^{-1},
\eea
where $\eta^{-1}$ is the inverse operator of $\eta$.

In general,  we would call an operator $O$ $\eta'$-pseudo-Hermitian for a Hermitian and invertible operator $\eta'$, if it satisfies $O^\dagger= \eta' O \eta'^{-1}$. Therefore, in the Hilbert space $\mathcal{H}_\eta$ it is more proper to take the $\eta$-pseudo-Hermitian operator $A$ as physical obverables. 

 Since $\eta$ is positive operator, we can define its ``square root'' $\eta^{1/2}$ which satisfies $(\eta^{1/2})^2=\eta$. It is obvious that
\bea\label{normrelation}
\langle \psi| \phi\rangle_\eta= \langle \eta^{1/2} \psi| \eta^{1/2} \phi\rangle. 
\eea
$\eta^{1/2}$ can be seen as the one-to-one maping between two Hilbert spaces.

Now consider the operators $a$ acting on $\mathcal{H}$. There is a corresponding operator $A$ acting on $\mathcal{H}_\eta$ given by
\bea
a=\eta^{-1/2} A \eta^{1/2},
\eea 
where $\eta^{-1/2}$ represents the inverse of the operator $\eta^{1/2}$. $a$ is Hermitian \textit{if and only if} $A$ satisfies
\bea\label{pseudohermitianoperator}
A^\dagger=\eta A \eta^{-1},
\eea
that is pseudo-Hermitian.

Now let us consider how to describe the state. In Hilbert space $\mathcal{H}$ one use the Hermitian density matrix to describe the system. Given a density matrix $\rho_\Psi=|\Psi\rangle \langle \Psi|$ in Hilbert space $\mathcal{H}_\eta$.  We have
\bea
\langle \Psi| A |\Psi\rangle_\eta=\langle \Psi| A \eta |\Psi\rangle= tr (A \mathcal{T}).
\eea 
Motivated by the definition  (\ref{densitymatrix}) in the Hilbert space $\mathcal{H}$ it is natural to introduce $\mathcal{T}:= \eta|\Psi\rangle\langle \Psi| $.
The non-Hermitian operator $\mathcal{T}$ in $\mathcal{H}$ plays the similar role as $\rho$ in $\mathcal{H}_\eta$. Since $\mathcal{H}_\eta$ and $\mathcal{H}$ are two equivalent spaces, we could make a conclusion that the non-Hermitian $\mathcal{T}$ can also be used as an equivalent way to describe the system as $\rho$.  
More generally, we can define the mixed state 
\bea
\rho_m= \sum_i p_i |\Psi_i\rangle\langle \Psi_i|,
\eea
where $p_i>0$ and $\sum_i p_i=1$. The expectation value of $A$ in the mixed state $\rho_m$ is
\bea
\sum_i p_i\langle \Psi_i |A |\Psi_i\rangle_\eta=\sum_i p_i\langle \Psi_i| A\eta |\Psi_i\rangle=tr(\eta\rho_m A).
\eea
This suggests to introcude the mixed transition matrix $\mathcal{T}_m:= \eta \rho_m$. 

In summary, if one uses the inner product (\ref{innerproductrelation}) associated with the metric operator $\eta$, we obtain a new definition of Hermitian conjugation. The observables in the Hilbert space $\mathcal{H}_\eta$ should be pseudo-Hermitian operators as defined in (\ref{pseudohermitianoperator}). The states can be more appropriately described using the transition matrix $\mathcal{T}$ (for pure states) and $\mathcal{T}_m$ (for mixed states). These are also pseudo-Hermitian operators, which can be easily verified by their definitions. Furthermore, there is a one-to-one correspondence between the transition matrix and the density matrix given by
\bea\label{hermitian_nonhermitian_dual}
\rho := \eta^{-1/2} \mathcal{T} \eta^{1/2},
\eea
which is consistent with (\ref{normrelation}). For mixed states, this correspondence also holds by replacing $\mathcal{T}$ with $\mathcal{T}_m$. We refer to this as \textit{Hermitian and non-Hermitian duality}.

\subsection{More general transition matrix}

In the above discussion, we have shown that the transition matrix can also be used to describe states, which naturally emerges when adapting different inner products. Specifically, in Euclidean QFTs, the transition matrix can be easily constructed.

In \cite{Nakata:2020luh}, the authors introduce the transition matrix $\trans := \frac{|\psi\rangle \langle \phi|}{\langle \phi|\psi\rangle}$ from a different motivation. When we use the Euclidean path integral to prepare states, one can prepare the bra $|\psi\rangle$ and the ket $\langle \phi|$ independently. If $|\phi\rangle \neq |\psi\rangle$, the path integral represents the transition matrix, which is non-Hermitian. Building on this idea, they define the reduced transition matrix of a subsystem $A$ 
\bea
\transA:= tr_{\bar A} \trans.
\eea
We can define the so-called pseudoentropy  $S(\transA):=-tr \transA \log \transA$ as a generalization of von Neumann entropy of $\rho_A:=tr\rho$. The primary motivation of \cite{Nakata:2020luh} is to generalize the RT formula to non-Hermitian cases with the assumption that the Euclidean path integral can be translated to bulk gravitational path integral via AdS/CFT correspondence.  

 The definition of the transition matrix in \cite{Nakata:2020luh} is quite general. One would expect that only a small set of transition matrices can be dual to a bulk geometry. According to the RT formula, in this case, the least constraint expected is that the pseudoentropy is positive. In \cite{Guo:2022jzs}, the authors propose a necessary and constructible condition within the framework of pseudo-Hermiticity to characterize the set of transition matrices that can be dual to a bulk geometry. However, the meaning of the pseudo-Hermitian condition is still unclear.

According to our previous discussion, we've observed that the pseudo-Hermitian condition is intimately linked to the inner product or quantization method of Euclidean QFTs. We demonstrated that the $\eta$-pseudo-Hermitian transition matrix emerges naturally when utilizing a new inner product associated with the metric operator $\eta$. In general, the class of transition matrices that are $\eta$-pseudo-Hermitian, with $\eta$ being positive, can also effectively describe the states of the system. Moreover, as indicated by relation (\ref{hermitian_nonhermitian_dual}), there exists a one-to-one correspondence between the $\eta$-pseudo-Hermitian transition matrix and the density matrix.

Let us consider an example in Euclidean QFTs. Using the basis $|\phi(x_0,x_1)\rangle$, the Hermitian conjugation is given by $\mathcal{O}(x_0,x_1)^\dagger =\mathcal{O}(-x_0,x_1)$. If one wants to consider the transition matrix 
\bea\label{transition_matrix_slice}
\mathcal{T}:= \mathcal{O}(-x_0,x_1)|0\rangle\langle 0|\mathcal{O}(x_0,x_1)e^{-2\tau_0 H},
\eea
which is non-Hermitian. It can be shown that $\mathcal{T}^\dagger=e^{-2\tau_0 H} \mathcal{T}e^{2\tau_0 H}$, thus $\mathcal{T}$ is $\eta$-pseudo-Hermitian with $\eta=e^{-2\tau_0 H}$. According to the relation (\ref{hermitian_nonhermitian_dual}) $\mathcal{T}$ corresponds to a Hermitian density matrix
\bea
\rho:= e^{-\tau_0 H} \mathcal{O}(-x_0,x_1)|0\rangle\langle 0|\mathcal{O}(x_0,x_1)e^{-\tau_0 H}.
\eea
According to the discussion in Section \ref{section_different_slice} $\rho$ can be understood as the density matrix with quantization of the theory on the timeslice $x_0=\tau_0$. $\rho$ and $\mathcal{T}$ describe the same state. There are also exact correspondence between the two descriptions. Specially, in next section we will consider the gravity dual of transition matrix in the context of AdS/CFT. The gravity dual of $\rho$ would help us to understand the details of the duality of $\mathcal{T}$. 

The above discussions can be generalized to arbitrary positive operators $\eta$. If one constructs an $\eta$-pseudo-Hermitian transition matrix $\mathcal{T}$ with $\eta$ being positive and invertible in Euclidean QFTs, it is possible to interpret it as a density matrix $\rho$ by using a different inner product where $\eta$ acts as the metric operator. Conversely, if one constructs a density matrix $\rho$, it can be understood as an $\eta$-pseudo-Hermitian transition matrix $\mathcal{T}$ by introducing a different inner product with $\eta$ as the metric. Denote $\mathcal{P}$ as the set of $\eta$-pseudo-Hermitian transition matrices $\mathcal{T}$ with $\eta$ being positive and invertible. The set $\mathcal{P}$ plays the same role as the set of density matrices. In the following sections, we will demonstrate that considering $\mathcal{P}$ is significant for understanding the AdS/CFT correspondence and the thermofield double formalism in general situations.

\section{Gravity dual of non-Hermitian transition matrices}\label{section_gravity_dual}

In previous sections, we focused on Euclidean QFTs and found it both natural and necessary to introduce the transition matrix to describe states in general cases. An interesting and important question is what kinds of transition matrices can be dual to a bulk geometry in the context of AdS/CFT.

If a transition matrix can be dual to bulk geometry, it should satisfy certain constraints, which arise from the entanglement quantities defined by the reduced transition matrix $\mathcal{T}_A$. From the perspective of QFTs, one could evaluate the pseudo-Rényi entropy in a process similar to that for the reduced density matrix\cite{Nakata:2020luh}. Thus, we expect that a bulk geometry dual of the pseudo-R\'enyi entropy exists if the transition matrix can be dual to bulk geometry as the holographic R\'enyi entropy \cite{Dong:2016fnf}\cite{Dong:2023bfy}. It is straightforward to generalize the holographic Rényi entropy formula to the transition matrix case. Therefore, the $n$-th pseudo-Rényi entropy should be positive. A necessary condition is that the spectrum of $\mathcal{T}_A$ should be positive.

In \cite{Guo:2022jzs}, it is shown that this necessary condition can be characterized by the pseudo-Hermitian condition, that is the transition matrix takes the form $$\trans=\frac{|\psi\rangle \langle \psi|\eta}{\langle \psi|\eta|\psi\rangle},$$ with $\eta$ being Hermtian and invertible operator. If $\eta$ can be factored as $\eta = \eta_A \otimes \eta_{\bar{A}}$ and both $\eta_A$ and $\eta_{\bar{A}}$ are positive or negative operators, it can be shown that the spectrum of $\mathcal{T}_A$ is positive. In fact, in this case, $\mathcal{T}_A$ can also be mapped to a density matrix. Some examples are demonstrated in \cite{Guo:2022jzs} \cite{Guo:2023tjv} to illustrate this idea.

As we showed in previous sections, the set $\mathcal{P}$ plays the same role as the set of density matrices. It is expected that the transition matrix in $\mathcal{P}$ can be dual to a bulk geometry in the semi-classical limit. We will focus on the pseudo-Hermitian transition matrix\footnote{In Section \ref{section_hermitian_nonhermitian}, the transition matrix is written as $\mathcal{T} = \eta |\Psi\rangle \langle \Psi|$. In the following, we will use the form (\ref{form2}), which is consistent with the notation used in \cite{Guo:2022jzs}. By replacing $|\Psi\rangle = \eta^{-1} |\Psi'\rangle$ in (\ref{form2}), we will obtain the form of $\mathcal{T}$ presented in Section \ref{section_hermitian_nonhermitian}. } 
\bea\label{form2}
\mathcal{T}=|\Psi\rangle \langle\Psi|\eta,
\eea
where $\eta$ is a positive operator. For mixed states, we can consider the transition matrix $\mathcal{T}_m = \sum_i p_i |\Psi_i\rangle \langle \Psi_i|\eta$. According to the discussions in previous sections, there exists a different inner product associated with the metric operator $\eta$, by which the non-Hermitian transition matrix is dual to a density matrix via the relation (\ref{hermitian_nonhermitian_dual}). Thus, if the density matrix has a well-defined dual bulk geometry, we expect $\mathcal{T}$ can also be dual to a geometry. In the following, we will use some examples to demonstrate the above idea and point out the relation of physical observables in the bulk. We refer to the metric that is dual to a non-Hermitian transition matrix as a non-Hermitian spacetime.

\subsection{Simple Examples}
Let us first explain how to understand the transition matrix on the bulk side, as it appears in Section \ref{section_different_time} and Section \ref{section_different_slice}.
\subsubsection{Bulk metric in AdS$_3$}
 We will focus on 2-dimensional CFTs. The dual bulk geometry can be constructed directly. Using the coordinate $w=x-i\tau$ and $\bar w=x+i\tau$, the general 3-dimensional AdS$_3$ can be written as the Banado geometry\cite{Banados:1998gg}
\bea\label{bulkgeometry}
ds^2=\frac{du^2+dw d\bar w}{u^2} +L(w)dw^2+\bar L(\bar w) d \bar w^2+ u^2 L(w)\bar L(\bar w) dw d\bar w,
\eea
where $L(w)=-\frac{6}{c}\langle T(w)\rangle$ and $\bar L(\bar w)=-\frac{6}{c} \langle \bar T(\bar w)\rangle$, the dual CFT lives on the boundary $u=0$, and $\langle T(w)\rangle$  and $\langle \bar T(\bar w)\rangle$ are expectation value of stress-energy tensor in the dual CFTs. Thus to obtain the geometry dual to transition matrix $\mathcal{T}$ we need to evaluate the expectation values $\langle T(w)\rangle_{\mathcal{T}}:=tr(\mathcal{T}T(w))$ and $\langle \bar T(\bar w)\rangle_{\mathcal{T}}:=tr(\mathcal{T}\bar T(\bar w))$.

In Poincaré coordinates, the AdS$_3$ metric is
\bea
ds^2=\frac{dz^2+d\xi d\bar \xi}{z^2},
\eea
$(\xi,\bar \xi)$ are the coordinates of the dual CFTs living on $z=0$. With the boundary conformal transformation $\xi=f(w)$ and $\bar \xi =\bar f(\bar w)$, the bulk coordinate transformation is given by
\bea\label{bulktransformation}
&&\xi=f(w)-\frac{2u^2f'(w)^2\bar f''(\bar w)}{4f'(w)\bar f'(\bar w)+u f''(w)\bar f''(\bar w)},\nn \\
&&\bar \xi=\bar f(\bar w)-\frac{2u^2\bar f'(\bar w)^2 f''( w)}{4\bar f'(\bar w) f'( w)+u \bar f''(\bar w) f''( w)},\nn \\
&&z=\frac{4u(f'(w)\bar f'(\bar w))^{3/2}}{4\bar f'(\bar w) f'( w)+u \bar f''(\bar w) f''( w)}.
\eea

The bulk metric in the coordinate $(u,w,\bar w)$ is given by (\ref{bulkgeometry}), with
\bea
&&L(w)=\frac{3f''(w)^2-2f'(w)f'''(w)}{4f'(w)^2},\nn\\
&&\bar L(\bar w)=\frac{3\bar f''(\bar w)^2-2\bar f'(\bar w)f'''(\bar w)}{4\bar f'(\bar w)^2}.
\eea
In the following sections we will use the above result to understand the bulk geometry dual to transition matrix and pseudoentropy.

\subsubsection{Geometry dual to (\ref{transition_matrix_time})}

In Section \ref{section_different_time}, we show that the state can be equivalently described by the density matrix (\ref{density_matrix_time}) and the transition matrix (\ref{transition_matrix_time}) with different choices of the ``time'' coordinate. For QFTs with a holographic dual, we can prepare the states $\rho$ and $\mathcal{T}$ via Euclidean path integral. This path integral can be translated to the gravity side through AdS/CFT. We can use this simple example to examine the bulk dual of the non-Hermitian transition matrix (\ref{transition_matrix_time}).

Given that the expectation values of observables in $\rho$ and $\mathcal{T}$ are equivalent, we anticipate them to yield the same spacetime metric $g_{\mu\nu}$. The expectation value of the stress-energy tensor $T(w)$ and $\bar{T}(\bar{w})$ would be the same for $\rho$ and $\mathcal{T}$. In AdS$_3$, the bulk metric is associated solely with these expectation values (\ref{bulkgeometry}). Thus, in the semiclassical limit $G \to 0$, the bulk geometry dual to $\rho$ and $\mathcal{T}$ would be the same. Specifically, we would have $tr\rho = tr\mathcal{T} \approx e^{-I_E}$, where $I_E$ denotes the Euclidean on-shell action of the specified bulk metric.

However, the metric $g_{\mu\nu}$ is invariant under the inversion $x_0 \to -x_0$ but not under $x_1 \to -x_1$. Consequently, we have two distinct perspectives on the same bulk geometry $g_{\mu\nu}$. If the boundary CFT is assumed to be described by the non-Hermitian matrix $\mathcal{T}$, $g_{\mu\nu}$ can be consistently regarded as a non-Hermitian spacetime. The path integral representation of $\mathcal{T}$ can be mapped to the bulk gravitational path integral using the AdS/CFT dictionary. Therefore, in the semiclassical limit $G \to 0$, a non-Hermitian geometry would dominate contributions to the Euclidean gravitational path integral.

The above arguments can be straightforwardly generalized to higher dimensions. In higher dimensions, near the AdS boundary in the Graham-Fefferman coordinate system \cite{Skenderis:2002wp}, the bulk metric $g_{\mu\nu}$ is related to the expectation values of the stress-energy tensor $T_{ij}$, which are the same for $\rho$ and $\mathcal{T}$. Thus, the bulk geometry would be the same, but the geometry has reflection symmetry along $x_0$ but not along other coordinates.

\subsubsection{Geometry dual to the transition matrix (\ref{transition_matrix_slice})}\label{section_geometry_slice}

In Section \ref{section_different_slice}, we introduce the transition matrix (\ref{transition_matrix_slice}). To simplify the notation, we choose $x_1=0$. The normalized transition matrix is
\bea\label{t2}
\mathcal{T}= \mathcal{N} \mathcal{O}(-x_0,0)|0\rangle \langle0| \mathcal{O}(x_0,0)e^{-2\tau_0 H},
\eea
where the normalization constant $\mathcal{N}=[4(x_0+\tau_0)^2]^{2h}$, $h$ is the conformal dimension of $\mathcal{O}$. We will take $\mathcal{O}$ to be a heavy operator, $h \sim O(c)$. Therefore, its backreaction on the geometry is significant. In this section, we explore its bulk geometry. With some calculations, we have
\bea\label{t2tt}
&&\langle T(w)\rangle_{\mathcal{T}}=-\frac{4h(x_0+\tau_0)^2}{[w^2+2i w \tau_0+x_0^2+2x_0 \tau_0]^2},\nn \\
&&\langle \bar T(\bar w)\rangle_{\mathcal{T}}=-\frac{4h(x_0+\tau_0)^2}{[\bar w^2-2i \bar w \tau_0+x_0^2+2x_0 \tau_0]^2}.
\eea
Taking the above expressions to (\ref{bulkgeometry}) we can obtain the bulk geometry associated with the transition matrix (\ref{t2}). It is obvious that $\langle T(w)\rangle_{\mathcal{T}}^*=\langle \bar T(\bar w)\rangle_{\mathcal{T}}$, where $*$ is the complex conjugation. Thus the metric (\ref{bulkgeometry}) is real using the coordinate $(x,\tau)$.

One could find a coordinate transformation mapping the coordinate (\ref{bulkgeometry}) with  (\ref{t2tt}) to Poincar\'e coordinate. The boundary conformal transformation is expected to be the form
\bea\label{conformal_mapping}
\xi=f(w)=\left(\frac{w-w_0}{w'_0-w}\right)^{\alpha_h},\quad \bar \xi=\bar f(\bar w)=\left(\frac{\bar w-\bar w_0}{\bar w'_0-\bar w}\right)^{\alpha_h}.
\eea
With some calculations we find
\bea
w'_0=i x_0, \ w_0=-i(x_0+2\tau_0),\ \alpha_h=\sqrt{1-24 h/c}.
\eea
One could obtain the bulk coordinate transformation by using (\ref{bulktransformation}). 

\subsection{Holographic pseudoentropy}

\begin{figure}
\centering\includegraphics[width=8cm]{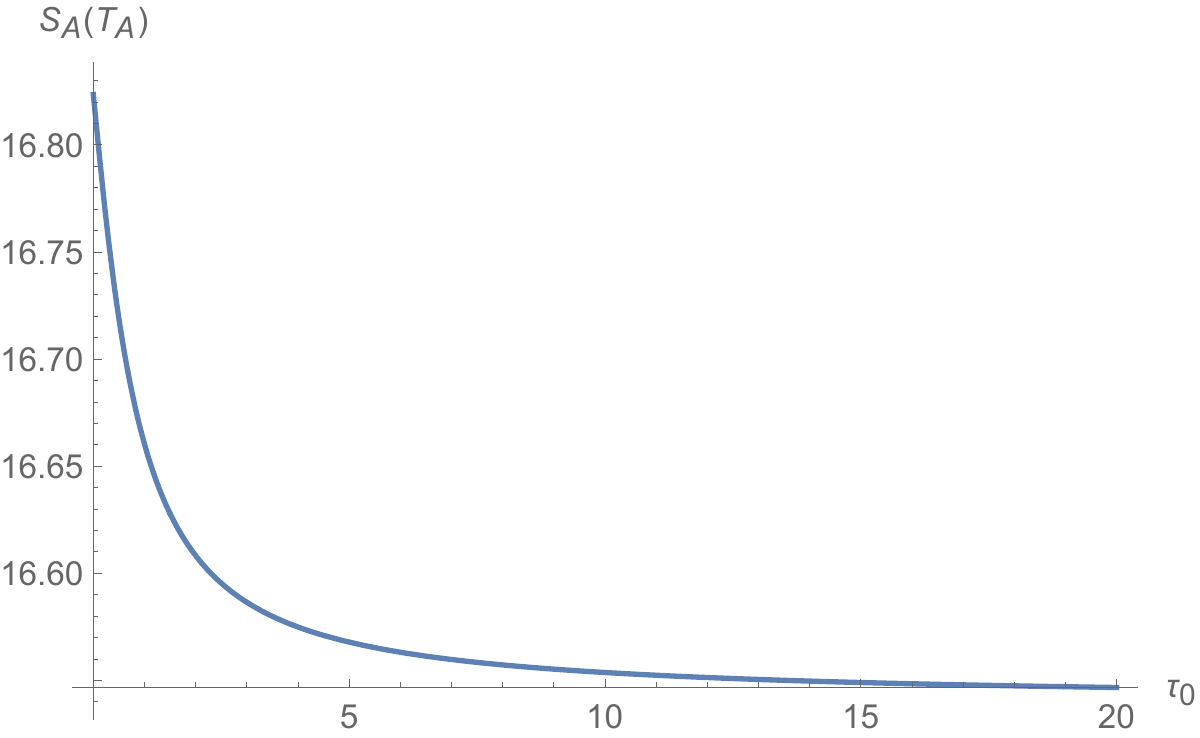}
\caption{Plot of $S(\mathcal{T}_A)$ for the interval $[0,R]$ as a function of $\tau_0$. We have chosed the parameters $R=2$, $x_0=1$, $\alpha_h=\frac{1}{3}$,$c=5$ and $\epsilon=10^{-4}$.}
\label{PEtau} 
\end{figure}

With the bulk metric one could evaluate the holographic pseudoentropy via the RT formula. Let us consider the metric in Section \ref{section_geometry_slice}. The subsystem $A$ is taken to be an interval $[0,R]$ on the timeslice $x_0=0$. The endpoints of $A$ are denoted by $(w_1,\bar w_1)=(0,0)$ and $(w_2,\bar w_2)=(R,R)$. We can define the reduced transition matrix $\mathcal{T}_A:=tr_{\bar A}\mathcal{T}$ and evaluate the pseudoentropy. One could obtain the minimal surface (line) by mapping to Poincar\'e 
coordinate. By the conformal transformation the two endpoints of $A$ are mapped to 
\bea\label{endpoints}
&&(\xi_1,\bar \xi_1)=(f(w_1),\bar f(\bar w_1))=\left((\frac{x_0+2\tau_0}{x_0})^{\alpha_h},(\frac{x_0+2\tau_0}{x_0})^{\alpha_h}\right),\nn\\
&&(\xi_2,\bar \xi_2)=(f(w_2),\bar f(\bar w_2))=\left((\frac{R+i (x_0+2\tau_0)}{i x_0-R})^{\alpha_h},(\frac{R-i (x_0+2\tau_0)}{-i x_0-R})^{\alpha_h}\right).\nn
\eea
 Using (\ref{conformal_mapping}), (\ref{endpoints}), and (\ref{SAgeneral}), it is straightforward to obtain the holographic pseudoentropy. For the expression of $S(\mathcal{T}_A)$ for a general interval $[a,b]$ on the timeslice $x_0=0$, one could refer to Appendix.\ref{section_holographic_pseudoentropy}. In the limit $\tau_0\to 0$, we obtain
\bea
\lim_{\tau_0\to 0}S(\mathcal{T}_A)=\frac{1}{6} c \log \left(\frac{\left(R^2+x_0^2\right) \left(-1+\left(-\frac{R-i x_0}{R+i x_0}\right)^{\alpha _h}\right) \left(-1+\left(-\frac{R+i x_0}{R-i x_0}\right)^{\alpha _h}\right)}{4 \epsilon ^2 \alpha _h^2}\right).
\eea
While in the limit $\tau_0\to \infty$ we have
\bea
\lim_{\tau_0\to \infty}S(\mathcal{T}_A)=\frac{1}{6} c \log \left(\frac{x_0^{1-\alpha _h} \left(R^2+x_0^2\right){}^{\frac{1}{2}-\frac{\alpha _h}{2}} \left(x_0^{\alpha _h}-\left(x_0-i R\right){}^{\alpha _h}\right) \left(x_0^{\alpha _h}-\left(x_0+i R\right){}^{\alpha _h}\right)}{\epsilon ^2 \alpha _h^2}\right).
\eea
It can be shown that the pseudoentropy for a fixed interval $[0,R]$ is a decreasing function of $\tau_0$. The results are ploted in Fig.\ref{PEtau}.

\begin{figure}
\centering\includegraphics[width=9cm]{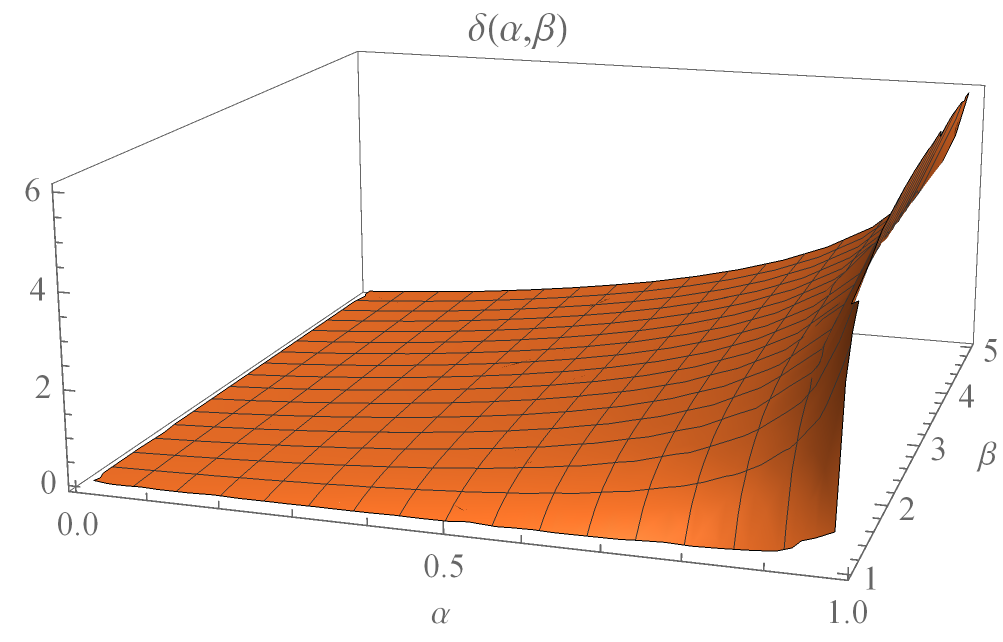}
\caption{Plot of $\delta(\alpha,\beta)$ (\ref{definition_delta}) as a function of $\alpha$ and $\beta$. We have chosen the parameters  $R=4$, $x_0=1$, $\tau_0=2$, $\alpha_h=\frac{1}{3}$,$c=5$ and $\epsilon=10^{-4}$.}
\label{SSA_plot} 
\end{figure}

The von Neumann entropy for density matrices satisfies the well-known property of strong subadditivity (SSA). Given  a density matrix $\rho_{ABC}$, we can define the  reduced density matrices $\rho_{AB}:=tr_{C}\rho_{ABC}$, $\rho_{BC}:=tr_{A}\rho_{ABC}$ and $\rho_B:=tr_{AC}\rho_{ABC}$. The von Neumann entropy for these density matrices the following  holds
\bea
S(\rho_{AB})+S(\rho_{BC})\ge S(\rho_{ABC})+S(\rho_B),
\eea
which is the SSA. Now we aim to verify whether SSA remains valid for the transition matrix. Our focus will be solely on the specific example discussed in this section. To achieve this, we select three intervals on the timeslice $x_0=0$,
\bea
A=[0,\alpha R],\quad B=[\alpha R,R],\quad C=[R,\beta R],
\eea
with $0<\alpha<1$ and $1<\beta<\infty$, and define 
\bea\label{definition_delta}
\delta(\alpha,\beta):=S(\mathcal{T}_{AB})+S(\mathcal{T}_{BC})-S(\mathcal{T}_{ABC})-S(\mathcal{T}_B).
\eea
The plot of $\delta(\alpha,\beta)$ in Fig.\ref{SSA_plot} illustrates that $\delta(\alpha,\beta)$ is always non-negative. For the general cases considered in this example, one can demonstrate the correctness of SSA. Further discussion on SSA is provided in Section \ref{section_discussion_SSA}.

\subsection{Observables correspondence}

In the previous section, we demonstrated the existence of a one-to-one correspondence between the transition matrix and the density matrix (\ref{hermitian_nonhermitian_dual}). Physical observables also exhibit certain relationships. For the transition matrix (\ref{t2}), the corresponding density matrix is given by
\bea\label{rho2}
\rho=\mathcal{N} e^{-\tau_0 H}\mathcal{O}(-x_0,0)|0\rangle \langle 0| \mathcal{O}(x_0,0)e^{-\tau_0 H},
\eea
where the normalization constant $\mathcal{N}$ is same as (\ref{t2}).
Let's examine the expectation values of the stress-energy tensor $T(w)$ and $\bar T(\bar w)$. We define
\bea
&&\langle T(w)\rangle_\mathcal{T}:= tr(T(w)\mathcal{T}),\quad \langle \bar T(\bar w)\rangle_\mathcal{T}:= tr(\bar T(\bar w)\mathcal{T}),\nn \\
&&\langle T(w)\rangle_\rho:= tr(T(w)\rho),\quad \langle \bar T(\bar w)\rangle_\mathcal{T}:= tr(\bar T(\bar w)\rho).
\eea
We have the following relation
\bea\label{stress_tensor_relation}
\langle T(w)\rangle_\mathcal{T}= \langle T(w+i\tau_0)\rangle_\rho,\quad \langle  \bar T(\bar w)\rangle_\mathcal{T}= \langle \bar T(\bar w-i\tau_0)\rangle_\rho,
\eea
where $T(w+i\tau_0):=e^{-\tau_0 H}T(w)e^{\tau_0H}$ and $\bar T(\bar w-i\tau_0):=e^{-\tau_0 H}\bar T(\bar w)e^{\tau_0H}$. One could also build the metric that is dual to $\rho$ (\ref{rho2}) by taking $\langle T(w)\rangle_\rho$ into (\ref{bulkgeometry}). In fact, the relation (\ref{stress_tensor_relation}) establishes the connections between two bulk geometries. 

Now let us consider the relation between holographic EE and PE. In QFTs one could evaluate the (pseudo) R\'enyi entropy by twist operators for $n$-copied CFT$_n$. For the transition matrix (\ref{t2}) and $A=[0,R]$ on the timeslice $x_0=0$, we have
\bea
tr(\mathcal{T}_A)^n= \langle \sigma_n(w_1,\bar w_1)\tilde{\sigma}_n(w_2,\bar w_2)\rangle_{\mathcal{T}^n},
\eea 
with $w_1=\bar w_1=0$ and $w_2=\bar w_2=R$, where $\mathcal{T}^n:=\mathcal{T}_{(1)}\otimes...\mathcal{T}_{(i)}\otimes ...\mathcal{T}_{(n)}$, the subsripts $i$ label the $i$-th copy. We can define an interval on the timeslice $x_0=\tau_0$ $A':=[i\tau_0,R+i\tau_0]$. By similar argument as the expectation value of $T(w)$ one could show 
\bea\label{trn_relation}
tr(\mathcal{T}_A)^n=\langle \sigma_n(w'_1,\bar w'_1)\tilde{\sigma}_n(w'_2,\bar w'_2)\rangle_{\rho^n}=tr (\rho_{A'})^n,
\eea
with $w'_1=i\tau_0$ and $w'_2=R+i\tau_0$. Thus, we obtain a relation between pseudo-R\'enyi entropy for $A$ in the state $\mathcal{T}$ and R\'enyi entropy for $A'$ in the state $\rho$. Using (\ref{trn_relation}) and definition of EE and PE it is straightforward to show $S(\mathcal{T}_A)=S(\rho_{A'})$, which is evident from the holographic perspective.

By the discussions in Section \ref{section_second} the density matrix $\rho$ and transition matrix $\mathcal{T}$ actually describe the same Euclidean path integral. It is natural that the expectation values of local operators are related. The above example is a special case with the metric operator $\eta=e^{-2\tau_0 H}$. For a general $\eta$, such a simple relation may not hold, but we expect that the above arguments can be generalized to more cases with appropriate modifications.

\subsection{Generalized thermofield double formalism and its holographic duality}

In previous sections we show the pseudo-Hermtian transition matrices $\mathcal{T}=|\Psi\rangle \langle \Psi|\eta$ and $\mathcal{T}_m:= \sum_i p_i |\Psi\rangle\langle \Psi|\eta $ can be used to describe the states of the system. Considering the holographic dual, the pseudo-Hermitian transition matrices provide additional insights into the states of gravity. These transition matrices have a one-to-one correspondence with the density matrix via the relation (\ref{hermitian_nonhermitian_dual}). In fact, one could use (\ref{hermitian_nonhermitian_dual}) to construct the dual bulk geometry for $\mathcal{T}$ once the duality for $\rho$ is established. We will demonstrate that the transition matrices are not only an equivalent way to describe the state but also essential for understanding the thermofield double formalism in general cases.

 %


\subsubsection{Thermofield double (TFD) state}
The thermofield double formalism is a trick to describe the thermal state $\rho=e^{-\beta H}$ with temperature $T=1/\beta$ by doubling the degree of freedom.  Consider two CFTs with Hamiltonian $H_1$ and $H_2$. These two theories live in different spacetime and have no interaction with each other. The unnormalized TFD state is defined as
 \bea\label{TFD_state}
 |\psi(\beta)\rangle=\sum_i e^{-\frac{\beta E_i}{2}}|i\rangle_1 |i\rangle_2, 
 \eea  
where $|i\rangle_1$, $|i\rangle_2$ are respectively the eigenstates of the Hamiltonian $H_1$ and $H_2$ of the two CFTs, $E_i$ are the eigenvalue. We have the nomralization $\langle \psi(\beta)|\psi(\beta)\rangle=Z(\beta):=\sum_i e^{-\beta  E_i}$, which is the partition function of thermal state with temperature $1/\beta$. $|\psi(\beta)\rangle$ is a pure state in the doubled Hilbert space $\mathcal{H}_1\otimes \mathcal{H}_2$. Its density matrix is
\bea\label{density_TFD}
\rho(\beta)=|\psi(\beta)\rangle\langle \psi(\beta)|.
\eea
The reduced density matrix of system $1(2)$ is given by
\bea\label{TFD1}
tr_{2(1)}\rho(\beta)=e^{-\beta H_{1(2)}},
\eea
Further, for any operators $\mathcal{O}_{1(2)}$ located in system $1(2)$ we would have
\bea\label{TFD2}
tr(\rho(\beta)\mathcal{O}_{1(2)})=\langle \psi(\beta)| \mathcal{O}_{1(2)}|\psi(\beta)\rangle=tr_{1(2)}(e^{-\beta H_{1(2)}}\mathcal{O}_{1(2)}).
\eea
The above two results (\ref{TFD1}) and (\ref{TFD2}) show the TFD state can be used to describe the thermal state for system $1(2)$ if we only focus on system $1(2)$. Even though there is no interaction between two systems, they are entangled with each other, that is there exists correlation between two systems. For operators $\mathcal{O}_{1(2)}$, generally $\langle \psi(\beta)| \mathcal{O}_1\mathcal{O}_2|\psi(\beta)\rangle \ne 0$. One could also evaluate the EE between two systems, which is equal to the thermal entropy for the thermal state $e^{-\beta H}$.

More interestingly, the TDF state can also be associated with eternal black hole in the framework of AdS/CFT. Consider the Hartle-Hawking-Israel state for the AdS eternal black hole with a temperature $T=1/\beta$. This state is defined by a path integral over half of the Euclidean section of the eternal black hole \cite{Israel:1976ur}. It has been argued in \cite{Maldacena:2001kr} that this state is dual to the thermofield double (TFD) state of two CFTs residing on the asymptotic boundary of AdS.  Thus, the density matrix (\ref{density_TFD}), which is expected to be dual to the gravitational path integral over the whole Euclidean section of eternal black hole. As a result we have $tr\rho(\beta)\approx e^{-I_E(\beta)}$ in the semi-classical limit $G\to 0$, where $I_E(\beta)$ is the on-shell action of the Euclidean black hole.

\subsubsection{Equivalent realizations of the thermofield  by transition matrix}

The TFD state is a way to purify the thermal state with introducing an identical auxiliary system. This is also known as canonical purification. There is  $1\leftrightarrow 2$ swap symmetry in the TFD state. In the following we would like to show there exists infinite way to realize the thermofield idea by using transition matrix while keeping the samp symmetry.

\begin{figure}
\centering\includegraphics[width=9cm]{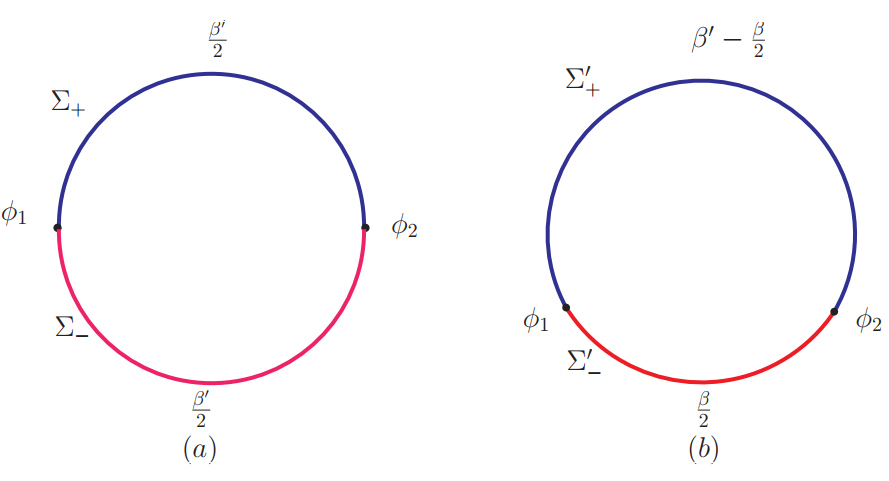}
\caption{Path integral representation of the density matrix $\rho(\beta')$ and $\mathcal{T}(\beta,\beta')$. The circle represents the Euclidean time coordinate. Each point in the diagram represents $S^{d-1}$. (a) illustrates $\rho(\beta')$. The boundary  with the field values $\phi_1$ and $\phi_2$ divides the circle with circumference $\beta'$ into two equal intervals, labeled by $\Sigma_-$(red) and $\Sigma_+$ (blue).  (b) illustrates $\mathcal{T}(\beta,\beta')$. The distinction from (a) lies in the placement of the boundary conditions $\phi_1$ and $\phi_2$. The circle is divided into two unequal intervals labeled by $\Sigma'_-$ (red) and $\Sigma'_+$ (blue). }
\label{f2} 
\end{figure}

Let's define the operator $H=H_1+H_2$. One can then explore the time evolution of the TFD state $e^{-iH t}|\psi(\beta)\rangle$. To further our investigation, we introduce the following unnormalized transition matrix,
\bea\label{generalized_TFD}
\mathcal{T}(\beta,\beta')=|\psi(\beta)\rangle \langle \psi(\beta)|e^{-\frac{\beta'-\beta}{2}H},
\eea
which is obviously pseudo-Hermitian. It can be related to a Hermitian density matrix $\rho(\beta')$ by a similarity transformation
\bea\label{TFD_correspondence}
\rho(\beta')= \eta^{1/2}\mathcal{T}(\beta,\beta')\eta^{-1/2},
\eea
where $\eta:=e^{-\frac{\beta'-\beta}{2}H}$. One could check $tr[\mathcal{T}(\beta,\beta')]= Z(\beta')$. According to previous discussion $\rho(\beta')$ is the density matrix of TFD state with temperature $T=1/\beta'$, which is expected to be dual to eternal black hole with temperature $1/\beta'$ in the gravity side.  We would now like to address and answer the following two questions:
\begin{itemize}
\item How can we interpret the non-Hermitian transition matrix $\mathcal{T}(\beta,\beta')$ in the CFTs?
\item In the semi-classical limit $G\to 0$, what is the bulk geometry $g_{\mu\nu}$ dual to $\mathcal{T}(\beta,\beta')$?
\end{itemize}
To understand the difference and relation between them, it is better for us to show the Eucldiean path integral representation. Assume the CFTs live on the sphere $S^{d-1}$. Specify the basis $|\phi_1\rangle$ and $|\phi_2\rangle$ for the two CFTs. 
The wave function $\langle \phi_1|\langle \phi_2|\psi(\beta')\rangle$ can  be prepared by Euclidean path integral over $\Sigma_-$ with boundary conditions $\phi_1$ and $\phi_2$ at the ends of the interval, where $\Sigma_-:=I_{\beta'/2}\times S^{d-1}$(see Fig.\ref{f2}). Similarly, $\langle \psi(\beta')|\phi_1\rangle |\phi_2\rangle$ is  given by path integral over $\Sigma_+$ shown in Fig.\ref{f2}. Thus the density matrix $\rho(\beta')$ can be understood as Euclidean path integral over the mainfold $\Sigma_-+\Sigma_+$ with specifying the field value $\phi_1$ and $\phi_2$ at the ends of $\Sigma_{\pm}$. 

The transition matrix $\mathcal{T}(\beta,\beta')$ can be prepared by Euclidean path integral similar to $\rho(\beta')$.   It can be represented as  path integral on $\Sigma'_-+\Sigma'_+$ shown in Fig.\ref{f2}, where $\Sigma'_-:=I_{\beta/2}\times S^{d-1}$ and $\Sigma'_+:= I_{\beta' -\beta/2}\times S^{d-1}$. The non-Hermitian property of $\mathcal{T}(\beta,\beta')$ manifests in the asymmetry between $\Sigma'_-$ and $\Sigma'_+$. 


In fact, $\rho(\beta')$ and $\mathcal{T}(\beta,\beta')$ can be viewed as equivalent realizations of the thermofield idea. The existence of equivalent ways to realize the thermofield idea beyond TFD state has been briefly discussed in \cite{Laflamme:1988wg}. It is straightforward to check that the reduced transition matrix of system $1(2)$  is just the density matrix of the system with temperature $1/\beta'$, 
\bea
tr_{2(1)} \mathcal{T}(\beta,\beta')=e^{-\beta' H_{1(2)}}.
\eea
Further, for arbitrary operators $\mathcal{O}_{1(2)}$ we have
\bea
tr_{1(2)}(\mathcal{O}_{1(2)}\mathcal{T}(\beta,\beta')) = tr (e^{-\beta' H_1} \mathcal{O}_{1(2)}).
\eea
It achieves the same effect as the TFD density matrix $\rho(\beta')$, see (\ref{TFD1}) and (\ref{TFD2}). 

Of course, $\rho(\beta')$ and $\mathcal{T}(\beta,\beta')$ are different by their definitions. $\rho(\beta')$ is constructed by considering two identical systems at $\tau=0$ and $\tau=\beta'/2$, whereas $\mathcal{T}(\beta,\beta')$ is constructed using the systems at $\tau=0$ and $\tau=\beta/2$. The positions of the two systems are different. One can observe that the correlator $tr [\mathcal{O}_1\mathcal{O}_2 \mathcal{T}(\beta,\beta')]$ differs from the one involving $\rho(\beta')$. By using (\ref{TFD_correspondence}) we could find the following correspondence
\bea\label{correlator_relation}
tr(\mathcal{O}_1 \mathcal{O}_2 \mathcal{T}(\beta,\beta'))=tr (\tilde{\mathcal{O}}_1 \tilde{\mathcal{O}}_2 \rho(\beta')),
\eea
where 
\bea
\tilde{\mathcal{O}}_{1(2)}:=e^{-\frac{\beta'-\beta}{4}H}\mathcal{O}_{1(2)} e^{\frac{\beta'-\beta}{4}H}.
\eea
 These properties become evident when we delve into the holographic dual in the subsequent discussion.

\subsubsection{Holographic explanation}\label{section_holographic_transition}

\begin{figure}
\centering\includegraphics[width=11cm]{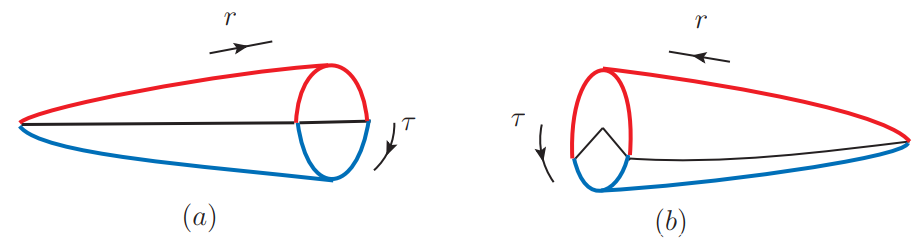}
\caption{Illustration of the bulk geometry with the sphere $S^{d-1}$ suppressed. $r$ is the holographic direction, $\tau$ is the Euclidean time. The CFTs lives on $r\to \infty$. The time circle shrinks to zero at the horizon of black hole. The circle represents the Euclidean time coordinate. (a) illustrates the gravity dual of $\rho(\beta')$. The region surrounded by the blue line can be regarded as dual to the region bounded by the boundary $\Sigma_-$, while the region surrounded by the red line is dual to $\Sigma_+$.  (b) illustrates the gravity dual of $\mathcal{T}(\beta,\beta')$. The regions surrounded by the blue line and red line correspond to the boundary $\Sigma'_-$ and $\Sigma'_+$ respectively.}
\label{f3} 
\end{figure}

The non-Hermitian transition matrix $\mathcal{T}(\beta,\beta')$ also finds a natural explanation within the framework of AdS/CFT. According to dictionary the boundary path integral can be translated to the bulk gravitational path integral. 

As we have mentioned the TFD state $|\psi(\beta')\rangle$ is expected to be dual to Hartle-Hawking-Israel state defined on half of the Euclidean black hole, see Fig.\ref{f3} for the illustration. The asymptotic boundary of the bulk solution ($r=\infty$) just gives the manifold $\Sigma_-$. The other half of the bulk solution is associated with its  Hermitian conjugation $\langle \psi(\beta')|$. Alternatively, we may say the density matrix $\rho(\beta')$ corresponds to the whole part of Euclidean black hole with temperature $1/\beta'$. We also have the partition function $Z(\beta')=\langle \psi(\beta')|\psi(\beta')\rangle\simeq e^{-I_{BH}}$ in the semi-classical limit $G\to 0$, where $I_{BH}$ is the on-shell action of the eternal black hole.

For the non-Hermitian $\mathcal{T}(\beta,\beta')$ we would also expect it should correspond to the 
same black hole solution. One evidence is that $tr\mathcal{T}(\beta,\beta')=tr\rho(\beta')=Z(\beta')$. From the perspective of the bulk, $\rho(\beta')$ and $\mathcal{T}(\beta,\beta')$ represent two distinct approaches to partitioning bulk spacetime, see Fig.\ref{f3} for the illustration. 

Further, both $\rho(\beta')$ and $\mathcal{T}(\beta,\beta')$ can be explained as two copies of the CFT in the entangled state. However, the locations of the two copies of CFT are different for the two cases. One could evaluate the entanglement between two CFTs using RT formula \cite{Ryu:2006bv}. Actually, for the transition matrix the RT formula would yield the pseudoentropy\cite{Nakata:2020luh}. Interestingly, the minimal surface remains the same for both cases, coinciding with the black hole horizon. Consequently, the entanglement entropy of the TFD density matrix $\rho(\beta')$ and pseudoentropy of the transition matrix $\mathcal{T}(\beta,\beta')$ are equal to the black hole entropy in both scenarios.  This is consistent with the fact that 
\bea
tr_{1(2)}\rho(\beta')=tr_{1(2)}\mathcal{T}(\beta,\beta').
\eea
The entropy would be equal for the two cases.
 However, the correlators involving $\mathcal{O}_1$ and $\mathcal{O}_2$ between the two CFTs exhibit differences in the two scenarios. From a holographic perspective, for very massive fields, the correlators can be approximated by geodesics passing through the black hole interior. It's evident from Fig.\ref{f3} that the geodesics would be distinct for the two cases. The relation (\ref{correlator_relation}) for the correlators also has a natural bulk explanation by the geodesic approximation.
 
Here we only consider the Eucldiean section of eternal black hole. It is still not clear how to link the Eucldiean part to Lorentizan solution of black hole. We will make comments on this issue in Section \ref{section_discussion}

Finally, let's summarize the answers to the two questions: The transition matrix $\mathcal{T}(\beta,\beta')$ serves as a broad manifestation of the thermofield idea. It corresponds to gravity states of eternal black hole with an asymmetrical partition. Moreover, $\mathcal{T}(\beta,\beta')$ can be interpreted as the states of two entangled CFTs on the asymptotic boundary of AdS. In this context, \textit{the black hole entropy emerges as  pseudoentropy.} This actually gives us a new way to understand black hole entropy. On the other hand it suggests the pseudoentropy can be taken as the real entropy, not pseudo one.

\subsection{Non-Hermitian spacetime in general cases}
\begin{figure}
\centering\includegraphics[width=8cm]{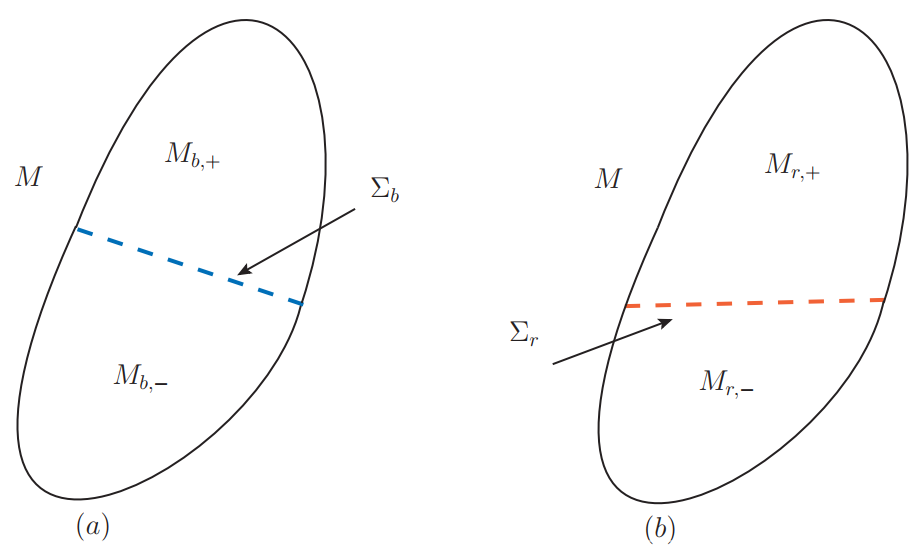}
\caption{Two possible partitions of the manifold $M$. Gravity states can be prepared using the Euclidean path integral. (a) depicts the symmetric partition, where the manifolds $M_{b,-}$ and $M_{b,+}$ can be mirrored onto each other by reflection across the submanifold $\Sigma_b$. (b) illustrates the asymmetric partition. }
\label{f4} 
\end{figure}
In previous discussions we have shown it is natrual to consider the pseudo-Hermitian transition matrix to describe the states. When consider Euclidean QFTs, different quantization may lead to different inner product or Hermitian conjugation operation. In the flat spacetime usually one could obtain the Euclidean QFTs via Wick rotation from the Lorentizian QFTs. In the eternal black hole example one could obtain the  Euclidean section of eternal black hole by Wick rotation since the spacetime is independent with time.  However, for general spacetime the Wick rotation may not work. 

To construct states of gravity using Euclidean gravitational path integral, one should  choose a codimension-1 submanifold $\Sigma$ with fixed induced metric $h_{ij}$ and matter field configuration $\phi_\Sigma$ \cite{Hartle:1983ai}. The submanifold $\Sigma$ divides the manifold $M$ into two parts, denoted by $M_-$ and $M_+$. With this division, one can prepare the gravity states by performing Euclidean path integrals over the two parts $M_{\pm}$. More precisely, the wave functions over $M_{\pm}$ are given by:
\bea\label{HHstate}
\Psi_{\pm}(h_{ij},\phi_\Sigma)=\int_{M_{\pm}}Dg_{\mu\nu}D\phi e^{-I_E},
\eea
where the path integral is over the possible $M$ with metrics $g_{\mu\nu}$ and matter field $\phi$ with the fixed boundary conditions on $\Sigma$. The product $P(h_{ij},\phi_\Sigma)=\Psi_{+}(h_{ij},\phi_\Sigma)\Psi_{-}(h_{ij},\phi_\Sigma)$ can be interpreted as the probability of obtaining the boundary values $h_{ij}$ and $\phi_\Sigma$ on $\Sigma$. 
It can also be understood as the expectation 
value in the basis  $|h_{ij},\phi_\Sigma\rangle$ on $\Sigma$, 
that is $\langle h_{ij},\phi_\Sigma|\hat \rho_M |h_{ij},\phi_\Sigma\rangle$, 
where $\hat\rho_M:=|\Psi_-\rangle\langle \Psi_+|$ is the density matrix associated with $M$.

However, there is no unique way to determine the position of the submanifold $\Sigma$. In Fig.\ref{f4}, we illustrate two potential choices denoted as $\Sigma_b$ and $\Sigma_r$. $M_{b,-}$ and $M_{b,+}$ exhibit reflection symmetry with respect to $\Sigma_b$, whereas $M_{r,-}$ and $M_{r,+}$ do not have this symmetry. In the case of symmetry, one could introduce the Euclidean ``time'' $\tau$ such that $\Sigma$ lies on the time slice $\tau=0$. The wave function $\Psi_{b,\pm}$, defined as (\ref{HHstate}) for $M_{b,\pm}$, will be the same. Alternatively, the density matrix $\hat\rho_M=|\Psi_{b,-}\rangle \langle \Psi_{b,+}|$ is Hermitian.  
In the non-symmetric case, $\Psi_{r,\pm}$ would indeed be different. This corresponds to the \textit{non-Hermitian} transition matrix denoted as $\hat \mathcal{T}_{\pm}:= |\Psi_{r,-}\rangle\langle \Psi_{r,+}|$. Moreover, in the general case for a given manifold $M$, there may not exist a submanifold $\Sigma$ that can divide $M$ into two equal parts. Therefore, considering non-Hermitian spacetime is thus inevitable in the gravitational path integral approach to quantum gravity. In the previous sections, we use some examples to show the above idea. For the eternal black hole case the significance of different partitions become evident.\\

\section{Conclusions and Discussions}\label{section_discussion}

In this paper, we have demonstrated the naturalness and necessity of considering non-Hermitian transition matrix and spacetime in the preparation of states in QFTs and quantum gravity via Euclidean path integral methods.

In general quantum systems, there is the freedom to choose the inner product, which is associated with the so-called metric operator $\eta$. We argue that one could also describe the state of the system using a transition matrix when employing a different inner product. The set of transition matrices, denoted by $\mathcal{P}$, is shown to be $\eta$-pseudo-Hermitian with $\eta$ being the metric operator. We propose that the set $\mathcal{P}$ can be used to describe the states of the system similarly to how density matrices are used. Moreover, there exists a one-to-one correspondence between the density matrices and the transition matrices (\ref{hermitian_nonhermitian_dual}).

Euclidean QFTs serve as good examples to illustrate the above idea. One could quantize Euclidean QFTs by choosing different time coordinates or different time slices. We use explicit examples to show how to describe the states using pseudo-Hermitian transition matrices. More interestingly, if the QFTs have a gravity dual, the Euclidean path integral in QFTs can be translated to bulk geometry via AdS/CFT. We refer to the geometry that is dual to the transition matrices as non-Hermitian spacetime. We provide examples to demonstrate the general properties of non-Hermitian spacetime.

There is an important question regarding which kinds of transition matrices can be dual to bulk geometry. We propose that $\eta$-pseudo-Hermitian transition matrices can have a bulk geometry dual with a real metric. These kinds of transition matrices have a one-to-one correspondence with density matrices. We expect that if one employs a different inner product with metric operator $\eta$, the $\eta$-pseudo-Hermitian transition matrices can be considered as density matrices in the new inner product space. Thus, if the density matrices can be dual to a bulk geometry, the corresponding transition matrices should also have a dual bulk geometry.

One may raise the question: Given that the $\eta$-pseudo-Hermitian transition matrix and density matrix are equivalent via the relation (\ref{hermitian_nonhermitian_dual}), what is the necessity for further investigation into the transition matrix? There are several reasons to consider the non-Hermitian transition matrix. As discussed in previous sections, quantizing the Euclidean QFTs is quite arbitrary. The transition matrix representation of a given state is necessary in general cases. Even though $\rho$ and $\mathcal{T}$ are equivalent, the reduced density matrix and transition matrix do exhibit some differences. This means the entanglement structure is different. Finally, we expect that the more general transition matrix $\mathcal{T}$ also has some physical meaning. In this general case, we do not expect there to be a correspondence to the density matrix.

We also construct the non-Hermitian spacetime using pseudo-Hermitian transition matrices, providing more opportunities to study AdS/CFT. Specifically, we show that non-Hermitian transition matrices are necessary to understand the thermofield double formalism, thus our results provide a generalization of the TFD density matrix. On the holographic aspect, the TFD states are proposed to be dual to the eternal black hole. The generalized non-Hermitian transition matrices (\ref{generalized_TFD}) are also proposed to be dual to the eternal black hole but with asymmetric partition of the spacetime as shown in Fig.\ref{f3}. We present some evidence for this proposal.

The non-Hermitian transition matrix and spacetime offer additional insights into the states of gravity. When using the Euclidean path integral to prepare the states of quantum gravity, it seems there exists different ways to define the inner product of the given states. Thus, transition matrices appear necessary to describe the states. Our results in this paper provide some examples to understand the states of gravity by using AdS/CFT. However, many interesting questions remain unsolved or untouched in this paper. Some of these questions are briefly discussed as follows.


\subsection{Strong subadditivity for transition matrix}\label{section_discussion_SSA}

SSA is a fundamental property of the von Neumann entropy of a density matrix. It is an intriguing question whether SSA applies to transition matrices. For general transition matrices, the pseudoentropy may be complex, making it meaningless to consider SSA. However, for a set of transition matrices with positive pseudoentropy, this becomes an interesting and unresolved problem.

As we discussed in previous sections we expect the $\eta$-pseudo-Hermitian transition matrix  with $\eta$ being positive operator may be the set with positive pseudoentropy. Let us consider the case $\eta$ can factor as $\eta_{A}\otimes \eta_{\bar A}$ for arbitrary $A$ with both $\eta_{A}$ and $\eta_{\bar A}$ being both positive. One could obtain 
\bea
\rho_{A}=\eta_A^{-1/2} \mathcal{T}_A \eta_{A}^{1/2},
\eea
where $\rho_{A}:=tr_{\bar A}\rho$ and $\mathcal{T}_A:= tr_{\bar A} \mathcal{T}$ \cite{Guo:2022jzs}. This means there exists non-Hermitian and Hermitian duality for the reduced matrices if $\eta=\eta_{A}\otimes \eta_{\bar A}$.

Now we consider three different subsystems $A,B$ and $C$. For the reduced transition matrix we would have
\bea
&&S(\mathcal{T}_{AB})+S(\mathcal{T}_{BC})-S(\mathcal{T}_{B})-S(\mathcal{T}_{ABC})\nn \\
&&=S(\eta_{AB}^{-1/2}\rho_{AB}\eta_{AB}^{1/2})+S(\eta_{BC}^{-1/2}\rho_{BC}\eta_{BC}^{1/2})-S(\eta_{B}^{-1/2}\rho_{B}\eta_{B}^{1/2})-S(\eta_{ABC}^{-1/2}\rho_{ABC}\eta_{ABC}^{1/2})\nn\\
&&=S(\rho_{AB})+S(\rho_{BC})-S(\rho_{B})-S(\rho_{ABC})\ge 0,
\eea
where we have used the entropy $S(\rho)$ is invariant under similarty transformation. Therefore, for the special case $\eta=\eta_{A}\otimes \eta_{\bar A}$ the SSA is also valid.
For the example (\ref{transition_matrix_slice}) we find it is $\eta$-pseudo-Hermitian with $\eta=e^{-\tau_0 H}$, which can be written as $e^{-\tau_0 H_A}\otimes e^{-\tau_0 H_{\bar A}}$ where $H_{A(\bar A)}=\int_{A(\bar A)} dx T_{00} $, $T_{00}$ is the energy density. In this example we do find the SSA is correct. 

However, it is nontivial to prove or disprove the SSA for more general transition matrix, especially the $\eta$-pseudo-Hermitian ones with $\eta\ne \eta_{ A}\otimes \eta_{\bar A}$. SSA is an important property for the von Neumann entropy of density matrix. For the set of transition matrices satisfying SSA, it is expected that the pseudoentropy can also be considered as a real entropy.


\subsection{Generalized thermofield formalism and eternal black hole}

In Section \ref{section_holographic_transition}, we explain the difference and relationship between $\rho(\beta')$ and $\mathcal{T}(\beta,\beta')$ from a holographic perspective. We propose that both may be dual to the same Euclidean eternal black hole but partitioned differently. Some evidence is provided to support this proposal. However, subtle questions remain regarding the gravitational states dual to $\mathcal{T}(\beta,\beta')$, especially when considering the Lorentzian spacetime.

The TFD state (\ref{TFD_state}) is prepared via the path integral over half of the Euclidean section. This preparation serves as the initial wavefunction, with the Lorentzian spacetime's half part obtained by evolving this initial wavefunction over time. The Euclidean and Lorentzian parts are joined at time $t=0$, where the spacetime exhibits time-reflection symmetry. The other half of the spacetime can be constructed similarly, representing the bra part of the TFD density matrix (\ref{density_TFD}).

In contrast, the transition matrix $\mathcal{T}(\beta,\beta')$ is associated with an asymmetrically cut Euclidean section of the eternal black hole, reflecting the non-Hermitian nature of the transition matrix. If we consider the time evolution of these initial states, it remains unclear what form the resulting Lorentzian spacetime would take. We will explore this problem in the near future.

\subsection{Complex metric and more general transition matrix}

In this paper, we primarily focus on Euclidean QFTs. Pseudo-Hermitian transition matrices can naturally appear when quantizing the theory using different methods. In these cases, we expect the bulk dual geometry to be real, as demonstrated in several examples. However, more general transition matrices may not be $\eta$-pseudo-Hermitian. If we assume these general transition matrices can also describe the state of the field theory and have a gravity dual, the corresponding gravity metrics would generally be complex.

In Euclidean QFTs, the examples we construct yield real metrics. However, if we consider the analytical continuation of Euclidean time to Lorentzian time, the transition matrices appear to lose their pseudo-Hermitian property, resulting in complex dual metrics.

Nonetheless, complex metrics seem necessary when considering the Euclidean path integral of gravity \cite{Gibbons:1976ue,Gibbons:1978ac,Halliwell:1989dy,Dong:2016hjy,Colin-Ellerin:2020mva,Saad:2018bqo}. Determining which complex metrics are allowable is an important and subtle problem \cite{Kontsevich:2021dmb,Witten:2021nzp}. As we have argued, for general transition matrices, the dual metrics are expected to be complex. It would be interesting to explore whether non-Hermitian transition matrices could aid in understanding and constructing these allowable complex metrics.

~
\\~
~\\
\textbf{Acknowledgements}

We would like to thank  Song He, Li Li, Rong-Xin Miao, Yu-Xuan Zhang for useful discussions. 
WZG is supposed by the National Natural Science Foundation of China under Grant No.12005070 and the Fundamental Research Funds for the Central Universities under Grants NO.2020kfyXJJS041.

\appendix

\section{Metric operators}\label{appendix_metric}

To define different inner product in finite dimensional Hilbert space, the metric operator $\eta$ plays the key role. The existence of the metric operator can be shown directly in finite dimensional Hilbert space. One could refer to \cite{Mostafazadeh:2004mx,Mostafazadeh:2008pw} for more discussions on the metric operator in the framework of pseudo-Hermitian quantum mechanics.

Let us firstly consider  a biothonoramal system, whose defintion is shown below. 
Consider $N$-dimensional Hilbert space $\mathcal{H}$. $\{|\mathfrak{e}_i\rangle\}$ ($i=1,...,N$) are the orthonormal and complete basis for $\mathcal{H}$. For any other basis $|\mathfrak{m}_i\rangle$ we have 
\bea
|\mathfrak{m}_i\rangle=\sum_j B_{ij} |\mathfrak{e}_j\rangle,
\eea
where $B_{ij}=\langle \mathfrak{e}_j |\mathfrak{m}_i\rangle$. The matrix $\{B_{ij}\}$ is invertible matrix. Let us define a new basis $\{\mathfrak{n}_i \}$ on the dual Hilbert space by using the matrix $\{B_{ij}\}$, 
\bea
\langle \mathfrak{n}_i| = \sum_i B^{-1}_{ji}\langle \mathfrak{e}_j|,
\eea
where $\{B^{-1}_{ij}\}$ is the inverse matrix of $\{ B_{ij}\}$. We can show $\langle \mathfrak{n}_i |\mathfrak{m}_j\rangle=\delta_{ij}$ and the completeness relation
\bea
\sum_i |\mathfrak{m}_i\rangle\langle \mathfrak{n}_i|=\sum_i |\mathfrak{n}_i\rangle\langle \mathfrak{m}_i|=\mathbb{I},
\eea
where $\mathbb{I}$ is the identity operator. The system is called biothonormal system if the Hilbert space $\mathcal{H}$ has the sequence $\{(|\mathfrak{m}_i\rangle, |\mathfrak{n}_i\rangle)\}$ satisfying the above relations. Now we would like to introduce the operator $\eta$
\bea
\eta:= \sum_i |\mathfrak{n}_i \rangle \langle \mathfrak{n}_i|,
\eea
which is Hermitian opertor. We can further show it is positive since for any state $|\psi\rangle$ we have $\langle \psi | \eta|\psi\rangle =\sum_i |\mathfrak{n}_i |\psi\rangle|^2 > 0$. It can be shown its inverse is given by
\bea
\eta^{-1}=\sum_i |\mathfrak{m}_i\rangle \langle \mathfrak{m}_i|. 
\eea
$\eta$ is the metric operator, which can be used to define a new and equivalent inner product of $\mathcal{H}$.

For arbitrary states $|\psi\rangle$ and $|\phi\rangle$ in $\mathcal{H}$, we can expand them in the basis $\{|\mathfrak{e}_i\}$ as
\bea
|\psi\rangle=\sum_i \psi_i |\mathfrak{e}_i \rangle,\quad |\phi\rangle=\sum_i \phi_i |\mathfrak{e}_i \rangle.
\eea
Usually, the inner product is defined as $\langle \psi|\phi\rangle=\sum_i \psi_i^* \phi_i$. One could also introduce a new inner product 
\bea
\langle \psi| \phi\rangle_\eta=\langle \psi|\eta\phi\rangle.
\eea
It is not hard to show it do satisfy the properties for innner product, such as $\langle \phi |\phi\rangle_\eta=\langle \phi| \eta |\phi\rangle \ge 0$.

Generally, for  a given vector space $\mathcal{V} $ we can introduce two different inner product $\langle\cdot |\cdot\rangle_1 $ and $\langle \cdot| \cdot \rangle_2$. Thus we would have two Hilbert spaces $\mathcal{H}_1$ and $\mathcal{H}_2$. Let $|\mathfrak{e}^{(1)}_i\rangle$ be an orthonormal  basis of $\mathcal{H}_1$, satisfying completeness relation $\sum_i |\mathfrak{e}^{(1)}_i\rangle \langle \mathfrak{e}^{(1)}_i|=\mathbb{I}$. In general, this set of states will not be an orthonormal basis of $\mathcal{H}_2$ since different inner product yields different Hermitian conjugation. In the Hilbert space $\mathcal{H}_2$ let us denote the basis as  $|\mathfrak{e}^{(2)}_i\rangle$ and define the operator
\bea\label{metric_finite_dimension}
\eta:= \sum_i |\mathfrak{e}^{(2)}_i\rangle \langle \mathfrak{e}^{(2)}_i|,
\eea
which is not the identity $\mathbb{I}$ in general. With this one could show the two inner products have the following relation,
\bea
\langle \psi| \phi\rangle_2=\langle \psi| \eta \phi\rangle_1,
\eea
where $\eta$ is the metric operator. Further, one could show the there is one-to-one correspondence between the metric operator and the inner product. 

\section{Details of the calculations of holographic pseudoentropy}\label{section_holographic_pseudoentropy}

In this section we would like to present some details of the calculations of holographic pseudoentropy for the transition matrix (\ref{t2}). Let us take the subsystem $A$ to be an interval $[a,b]$ on the timeslice $x_0=0$. The endpoints of $A$ are $(w_1,\bar w_1)=(a,a)$ and $(w_2,\bar w_2)=(b,b)$. By the conformal transformation (\ref{conformal_mapping}) they are mapped to the  complex $\xi$ plane with the coordinates
\bea\label{endpoints_appendix}
&&(\xi_1,\bar \xi_1)=(f(w_1),\bar f(\bar w_1))=\left(\left(\frac{a+i \left(2 \tau _0+x_0\right)}{-a+i x_0}\right)^{\alpha _h},\left(\frac{a-i \left(2 \tau _0+x_0\right)}{-a-i x_0}\right)^{\alpha _h}\right),\nn\\
&&(\xi_2,\bar \xi_2)=(f(w_2),\bar f(\bar w_2))=\left(\left(\frac{b+i \left(2 \tau _0+x_0\right)}{-b+i x_0}\right)^{\alpha _h},\left(\frac{b-i \left(2 \tau _0+x_0\right)}{-b-i x_0}\right)^{\alpha _h}\right).\nn
\eea
To evaluate holographic pseudoentropy one could calculate the geodesic line connecting two points $(\xi_1,\bar \xi_1)$ and $(\xi_2,\bar \xi_2)$ in the Poincare coordinate. 
 By using the coordinate transformation (\ref{bulktransformation}) with $f(w)$ and $\bar f(\bar w)$ being (\ref{conformal_mapping}), one could obtain the geodesic line in the metric dual to (\ref{t2}). 
The holographic pseudoentropy is given by
\bea\label{holographic_pseudo}
S(\mathcal{T}_A)=\frac{c}{6}\log \frac{(f(w_1)-f(w_2))(\bar f(\bar w_1)-\bar f(\bar w_2))}{\epsilon^2 \sqrt{f'(w_1)\bar f'(\bar w_1)f'(w_2)\bar f'(\bar w_2)}},
\eea
where $\epsilon$ is the UV cut-off.
Taking (\ref{endpoints_appendix}) into (\ref{holographic_pseudo}) we have
\bea\label{SAgeneral}
&&S(\mathcal{T}_A) \nn\\
&&=\frac{1}{6} c \log \left[\frac{\left(\left(\frac{-a+i \left(2 \tau _0+x_0\right)}{a+i x_0}\right)^{\alpha _h}-\left(\frac{-b+i \left(2 \tau _0+x_0\right)}{b+i x_0}\right)^{\alpha _h}\right) \left(\left(\frac{a+i \left(2 \tau _0+x_0\right)}{-a+i x_0}\right)^{\alpha _h}-\left(\frac{b+i \left(2 \tau _0+x_0\right)}{-b+i x_0}\right)^{\alpha _h}\right)}{4 \epsilon^2 \alpha _h^2 \left(\tau _0+x_0\right)^2}\right]\nn \\
&&\quad+\frac{1}{6} c \log \left[\frac{\left(a^2+x_0^2\right)^{\frac{1}{2} \left(\alpha _h+1\right)} \left(b^2+x_0^2\right)^{\frac{1}{2} \left(\alpha _h+1\right)}}{\left(a^2+\left(2 \tau _0+x_0\right)^2\right)^{\frac{1}{2} \left(\alpha _h-1\right)} \left(b^2+\left(2 \tau _0+x_0\right)^2\right){}^{\frac{1}{2} \left(\alpha _h-1\right)}}\right].
\eea
Using the above result we can discuss the properties of holographic pseudoentropy. 


\end{document}